\documentclass[%
 aip,
 amsmath,amssymb,
 reprint,%
longbibliography
]{revtex4-1}

\usepackage{graphicx}
\usepackage{dcolumn}
\usepackage{bm}

\usepackage[T1]{fontenc}
\usepackage{mathptmx}
\usepackage{etoolbox}
\usepackage[abs]{overpic}
\usepackage{xcolor}
\usepackage{ulem,comment}
\makeatletter
\def\@email#1#2{%
 \endgroup
 \patchcmd{\titleblock@produce}
  {\frontmatter@RRAPformat}
  {\frontmatter@RRAPformat{\produce@RRAP{*#1\href{mailto:#2}{#2}}}\frontmatter@RRAPformat}
  {}{}
}%
\makeatother
\begin{document}

\preprint{AIP/123-QED}

\title[Generation-Resolved Signatures in QED Cascades: Diagnostics for Ultraintense Laser Parameters]{Generation-Resolved  Signatures in QED Cascades: Diagnostics for Ultraintense Laser Parameters}
\author{Ke-Jia Wei}
\affiliation{Ministry of Education Key Laboratory for Nonequilibrium Synthesis and Modulation of Condensed Matter, State key laboratory of electrical insulation and power equipment, Shaanxi Province Key Laboratory of Quantum Information and Quantum Optoelectronic Devices, School of Physics, Xi'an Jiaotong University, Xi'an 710049, China}
\author{Feng Wan}\email{wanfeng@xjtu.edu.cn}
\affiliation{Ministry of Education Key Laboratory for Nonequilibrium Synthesis and Modulation of Condensed Matter, State key laboratory of electrical insulation and power equipment, Shaanxi Province Key Laboratory of Quantum Information and Quantum Optoelectronic Devices, School of Physics, Xi'an Jiaotong University, Xi'an 710049, China}
\author{Hao-Tian Wang }
\affiliation{Ministry of Education Key Laboratory for Nonequilibrium Synthesis and Modulation of Condensed Matter, State key laboratory of electrical insulation and power equipment, Shaanxi Province Key Laboratory of Quantum Information and Quantum Optoelectronic Devices, School of Physics, Xi'an Jiaotong University, Xi'an 710049, China}%
\author{Yan-Xi Wu}
\affiliation{Ministry of Education Key Laboratory for Nonequilibrium Synthesis and Modulation of Condensed Matter, State key laboratory of electrical insulation and power equipment, Shaanxi Province Key Laboratory of Quantum Information and Quantum Optoelectronic Devices, School of Physics, Xi'an Jiaotong University, Xi'an 710049, China}%
\author{He Yuan}
\affiliation{Ministry of Education Key Laboratory for Nonequilibrium Synthesis and Modulation of Condensed Matter, State key laboratory of electrical insulation and power equipment, Shaanxi Province Key Laboratory of Quantum Information and Quantum Optoelectronic Devices, School of Physics, Xi'an Jiaotong University, Xi'an 710049, China}%
\author{Jian-Xing Li}\email{jianxing@xjtu.edu.cn}
	\affiliation{Ministry of Education Key Laboratory for Nonequilibrium Synthesis and Modulation of Condensed Matter, State key laboratory of electrical insulation and power equipment, Shaanxi Province Key Laboratory
    of Quantum Information and Quantum Optoelectronic Devices, School of Physics, Xi'an Jiaotong University, Xi'an 710049, China}
	\affiliation{Department of Nuclear Physics, China Institute of Atomic Energy, P. O. Box 275(7), Beijing 102413, China}
\date{\today}

\begin{abstract}
We present a generation-resolved analysis of shower-type, spin- and polarization-dependent quantum electrodynamics (QED) cascades initiated by head-on collisions of ultraintense laser pulses ($a_0 = 200$--$1000$) with $10$~GeV electron bunches. Using a Monte Carlo model that tracks cascade evolution across distinct generations, we shows that radiation reaction strongly suppresses high-generation yields. The positron energy spectrum exhibits a systematic softening with increasing $a_0$, and the average photon polarization $\overline{\xi}_3$ increases with pulse duration $\tau$ due to polarization-selective depletion in nonlinear Breit--Wheeler pair production. We identify a scaling relation $a_0^2\tau$ that governs both the maximum cascade generation $G_{\max}$ and the fraction of backward-emitted photons $F_r$, thereby establishing experimentally accessible diagnostics for laser intensity and pulse duration using two observables: $F_r$ and the positron peak energy $\varepsilon_+^{\rm peak}$. These generation-dependent signatures suggest a potential shot-resolved, post-interaction approach for characterization of ultraintense laser parameters in the strong-field QED regime.
\end{abstract}

\maketitle

\section{\label{sec:level1}Introduction}

When matter or particles interact with ultra-intense electromagnetic fields, quantum electrodynamics (QED) cascades driven by nonlinear Compton scattering (NCS) and nonlinear Breit-Wheeler (NBW) pair production~\cite{DiPiazza2012,Bell2008,Bulanov2013,Fedotov2023}, play a key role in extreme nonlinear phenomena in strong-field physics. 
These cascades operate in two distinct regimes: the avalanche type, characterized by exponential plasma growth driven by self-sustained particle acceleration in a standing-wave or counter-propagating laser geometry~\cite{Fedotov2010,Grismayer2016, Seipt2021}, and the shower-type, initiated by high-energy seed particles traversing a single intense laser pulse~\cite{Gonoskov2022,Mironov2014,Bulanov2013,Pouyez2025}. The present work focuses exclusively on the shower regime.

In extreme astrophysical environments---such as black hole magnetospheres~\cite{Blandford1977,Ford2018}, pulsar polar caps~\cite{Goldreich1969,Medin2010,Timokhin2019,Cruz2021,Song2024}, and gamma-ray bursts~\cite{Piran2004,Kumar2015,Ravasio2024}---these processes govern the formation of relativistic electron-positron jets and the mechanisms of energy conversion. 
In laboratory strong-field physics, cascades impose a theoretical limit on attainable laser intensities due to rapid field depletion~\cite{Fedotov2010,Gonoskov2022}. Cascades may also serve as the potentially primary mechanism for generating macroscopic, quasi-neutral, dense electron-positron pair plasmas.
Such pair plasmas offer unique platforms to study collective QED behaviors, which are difficult in conventional electron-ion plasmas~\cite{Arrowsmith2024,Sarri2025,DiPiazza2012}. 
Furthermore, cascades play a key role in particle polarization effects in extreme fields~\cite{Li2019,Seipt2020,Seipt2021}.
Yet experimental investigation remains elusive due to insufficient laser intensities to trigger these processes~\cite{Fedotov2010,Gonoskov2022,Qu2024POP}.

Recent advances in ultra-intense laser technology have enabled laboratory studies of QED cascades. 
Facilities such as CoReLS~\cite{Yoon2021}, SULF~\cite{Li2022SULF}, ELI-NP~\cite{Radier2022}, and Vulcan~\cite{stfcVulcan2020} are operational or under development toward intensities of $10^{22}$--$10^{23}~\mathrm{W/cm^2}$, with CoReLS already exceeding $10^{23}~\mathrm{W/cm^2}$ and 100-PW-class facilities such as XCELS~\cite{xcels,Khazanov2023} anticipated in the near future.
These developments provide the experimental basis for investigating QED cascade dynamics with tunable laser parameters. Moreover, the rapid development of laser--plasma acceleration has made high-quality multi-GeV electron beams increasingly available, further supporting the laboratory realization of shower-type QED cascades~\cite{Zhu2023,Babjak2024,Tsymbalov2025}.

\begin{figure*}
    \centering
    \begin{overpic}{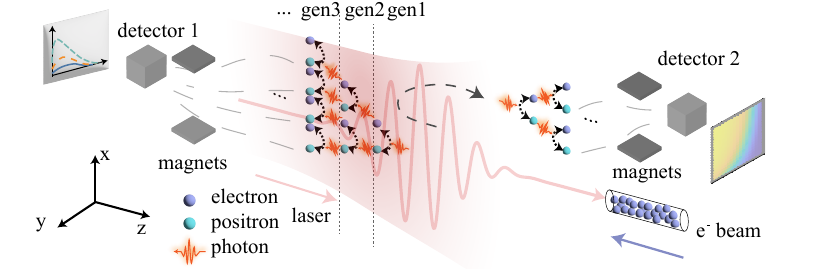}\label{fig1}
    \end{overpic}
    \caption{Schematic of a laser--electron interaction producing a generation-resolved QED cascade and its diagnostics. A relativistic electron bunch injected from $-z$ enters the intense laser focal region (red), emitting high-energy photons (orange) via nonlinear Compton scattering. These photons subsequently convert into electron--positron pairs through the NBW process, thereby forming a multi-step cascade. Detector~1 counts the forward-going photons and measures the positron energy distribution, whereas Detector~2 records photons and positrons emitted in the opposite direction. From the fraction of backward-emitted photons and/or the positron energy spectrum, one can infer the number of cascade generations and diagnose the laser parameters.}
\end{figure*}

Paradoxically, while the availability of ultra-intense lasers enables QED cascade experiments, the precise characterization of such laser parameters becomes extraordinarily challenging. Conventional diagnostic techniques are no longer reliable as the nonlinear QED effects dominate the interaction. 
Standard optical reconstruction methods infer the peak intensity from separate measurements of the pulse energy, temporal profile, and focal-spot distribution, with accumulated uncertainties and full-power aberrations potentially leading to discrepancies of up to approximately $50\%$~\cite{Trebino1997,Alonso2024,Lu2023}.
Atomic-physics-based diagnostics, which leverage tunneling ionization~\cite{Smeenk2011,Ciappina2019,Ciappina2020} or photoelectron yields~\cite{Pullen2013,Zhao2016,Aleksandrov2021}, become inapplicable because target atoms undergo complete barrier-suppression ionization long before the pulse reaches the peak QED intensity. 
Similarly, free-electron-based diagnostics relying on electron energy or angular distributions~\cite{Gao2006,Vais2017,Ravichandran2023,HarShemesh2012,Xie2025} become increasingly difficult to interpret in the QED-cascade regime, where radiation reaction and successive stochastic emissions reshape the primary-electron signal and redistribute its energy among secondary photons and pairs~\cite{HarShemesh2012,Bulanov2013,Pouyez2025}.
Consequently, uncertainties in the laser parameters, particularly the peak intensity, which strongly influences the cascade growth rate, lead to uncertainties in predicted cascade observables and their experimental interpretation. Accurate in situ determination of the laser parameters is therefore essential for quantitative tests of strong-field QED.

To address this critical challenge, we focus exclusively on the shower regime and investigate spin- and polarization-resolved QED cascades in a laser-electron interaction configuration. Our simulations demonstrate that laser parameters significantly imprint their signatures on key cascade observables, specifically the positron energy spectra, the polarization degree of emitted photons, and their momentum directions. These distinct correlations reveal a fundamental insight: the statistical properties of QED cascade products carry embedded, quantitative information about the driving laser field. Exploiting this  dependence, we propose a novel diagnostic scheme that enables  the retrieval of laser intensities and pulse durations from cascade signatures. This approach transforms the QED cascade itself from a subject of study into a precise, self-consistent diagnostic tool for strong-field conditions, thereby closing the loop between laser parameter uncertainty and cascade measurement precision.

\section{Generation-Resolved Monte Carlo Framework}
\label{sec2}

We consider a shower-type QED cascade initiated by the head-on collision of a relativistic electron bunch with an ultra-intense laser pulse. Throughout this work, natural units with $\hbar=c=1$ are used.

The laser pulse is characterized by two control parameters: the normalized peak field amplitude
\begin{equation}
a_0=\frac{eE_0}{m\omega},
\end{equation}
and the pulse duration $\tau$, where $-e$ and $m$ are the electron charge and rest mass, respectively, and $E_0$ and $\omega$ are the peak electric-field amplitude and angular frequency of the laser. We scan the ranges
\begin{equation}
200\leq a_0\leq 1000,
\qquad
2T_0\leq\tau\leq12T_0,
\end{equation}
where $T_0=2\pi/\omega$ is the laser period. All other laser and electron-bunch parameters are held fixed and are specified in Sec.~\ref{sec3}. The selected parameter window covers the regime targeted by the present cascade-based diagnostic. At lower amplitudes, established ultrashort-pulse characterization and electron-based diagnostic methods remain applicable~\cite{Trebino1997,Alonso2024,Lu2023}, whereas at substantially higher amplitudes the approximations adopted in the present model, including the prescribed-field and collinear-emission treatments, require further assessment~\cite{DiPiazza2019,Montefiori2023}.

The strength of the strong-field QED interaction is characterized by the quantum nonlinearity parameters
\begin{equation}
\chi_e
=
\frac{e}{m^3}
\sqrt{-\left(F_{\mu\nu}p^\nu\right)^2},
\qquad
\chi_\gamma
=
\frac{e}{m^3}
\sqrt{-\left(F_{\mu\nu}k^\nu\right)^2},
\end{equation}
where $F_{\mu\nu}$ is the electromagnetic field tensor, and $p^\nu$ and $k^\nu$ are the four-momenta of a charged particle and a photon, respectively~\cite{Baier1973}.

We employ the local constant field approximation (LCFA), under which the formation length of each strong-field QED process is assumed to be much shorter than the characteristic scale over which the background field varies. For nonlinear Compton scattering in a plane-wave background, an energy-resolved LCFA validity condition can be written as~\cite{DiPiazza2018,DiPiazza2019}
\begin{equation}
\eta_{\mathrm{LCFA}}
=
\frac{1-s}{s}
\frac{\chi_e}{a_0^3}
\ll 1,
\qquad
s=\frac{k_-}{p_-},
\end{equation}
where $s$ is the fraction of the initial electron light-front momentum carried by the emitted photon. For photons carrying a non-negligible fraction of the electron momentum, this condition reduces approximately to
\begin{equation}
\frac{\chi_e}{a_0^3}\ll1.
\end{equation}
Over the parameter range considered here, the maximum value of this ratio is approximately
\begin{equation}
\frac{\chi_e}{a_0^3}\simeq2\times10^{-6}.
\end{equation}
The LCFA is therefore well justified.

Within the LCFA, the angle-integrated differential probability for NCS, including electron-spin and photon-polarization effects, is written as~\cite{Li2019,Baier1973}
\begin{equation}
\label{NCS}
\frac{\mathrm{d}^2W_{\mathrm{NCS}}}
{\mathrm{d}u,\mathrm{d}\eta}
=
\frac{W_R}{2}
\left(
F_0+\xi_1F_1+\xi_2F_2+\xi_3F_3
\right).
\end{equation}
The corresponding differential probability for nonlinear Breit--Wheeler pair production is
\begin{equation}
\label{NBW}
\frac{\mathrm{d}^2W_{\mathrm{NBW}}}
{\mathrm{d}u,\mathrm{d}\eta}
=
\frac{1}{2}
\left(
G_0+\tilde{\xi}_1G_1
+\tilde{\xi}_2G_2
+\tilde{\xi}_3G_3
\right).
\end{equation}
Here, $\eta$ denotes the laser phase, $u$ is the process-dependent energy-sharing variable, and $W_R$ is the common prefactor of the nonlinear Compton rate. 
The photon polarization is described by the Stokes vector $\bm{\xi}=(\xi_1,\xi_2,\xi_3)$. Here, $\xi_1$ characterizes linear polarization along axes rotated by $\pm45^\circ$ relative to the chosen transverse basis, $\xi_2$ describes circular polarization, and $\xi_3$ represents the difference between the two orthogonal linear-polarization components. The quantities $\tilde{\xi}_i$ denote the same polarization state transformed into the local basis used for the NBW process.
The explicit expressions for $F_i$, $G_i$, and the spin- and polarization-resolved transition probabilities are given in Refs.~\cite{Li2019,Baier1973,Xue2022}.

To resolve the cascade hierarchy, we use the following generation-labeling convention. The initial seed electrons constitute generation $0$ (gen-$0$). Photons emitted by charged particles of generation $n$ (gen-$n$) are assigned to the same generation, while electron--positron pairs produced by these photons are assigned to generation $n+1$ (gen-$(n+1)$). The charged-particle generation number therefore corresponds directly to the number of NBW pair-production steps.

The cascade is simulated using the iterative Monte Carlo procedure developed in Ref.~\cite{Xue2022}, which is applicable to the quantum radiation-dominated regime under dilute-beam conditions.
Between quantum events, charged particles are propagated in the external laser field according to the Lorentz equation, while their spin evolution is described by the Thomas--Bargmann--Michel--Telegdi equation~\cite{Thomas1926,Thomas1927,Bargmann1959}. Photons propagate ballistically until they leave the interaction region or undergo pair production.

At each time step, NCS and NBW events are sampled from the initial-state-resolved differential rates
\begin{equation}
\frac{\mathrm{d}^2W}
{\mathrm{d}u,\mathrm{d}\eta}
\left(S_i,p_i\right)
\end{equation}
using the von Neumann rejection method. Once an event is accepted, the final momentum, spin, and polarization states are sampled from the fully resolved transition probabilities
\begin{equation}
\frac{\mathrm{d}^2W}
{\mathrm{d}u,\mathrm{d}\eta}
\left(S_i,S_f,p_i,p_f\right),
\end{equation}
where $S_i$ and $S_f$ denote the initial and final spin or polarization states, and $p_i$ and $p_f$ denote the corresponding initial and final momenta~\cite{Xue2022}.

All particles generated during iteration $n$ are stored together with their phase-space coordinates and quantum states and are subsequently used as the input for iteration $n+1$. We define the maximum cascade generation, $G_{\max}$, as the highest generation whose particle yield remains at or above $1\%$ of the initial seed-electron population. The iterative procedure is terminated once the yield of the newly generated particles falls below this threshold. Numerical convergence tests confirm that the resulting values of $G_{\max}$, particle yields, and spectral observables are stable with respect to the number of sampled seed electrons.

\section{Shower-Type QED Cascade Dynamics}
\label{sec3}

The interaction geometry is illustrated in Fig.~\ref{fig1}. We first examine the representative case $a_0=600$ and $\tau=10T_0$, while the remaining parameter combinations are obtained using the scan specified in Sec.~\ref{sec2}. The laser pulse is linearly polarized along the $x$ direction and propagates along the $+z$ direction. It has a wavelength of $\lambda=1~\mu\mathrm{m}$ and a focal waist of $w_0=5~\mu\mathrm{m}$. For these parameters, $a_0=600$ corresponds to a peak intensity of approximately $5\times10^{23}~\mathrm{W/cm^2}$.

\begin{figure}[!h]
    \centering
    \begin{overpic}{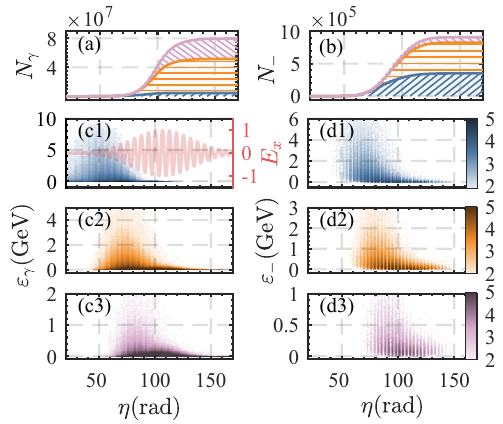}
    \end{overpic}
     \caption{\label{fig2}(a) and (b) Generation-resolved cumulative yields of photons $N_\gamma$ and electrons $N_-$ vs phase $\eta$: blue line $N$(gen-1), yellow line $N$(gen-1, gen-2), and lavender line $N$(gen-1, gen-2, gen-3). Phase and energy-resolved photon number density $\log_{10}\left[\mathrm{d}^2 N / (\mathrm{d}\eta\mathrm{d}\varepsilon)~(\mathrm{GeV}^{-1})\right]$ for photons (c1--c3) and electrons (d1--d3), and the second, third, and fourth rows stand for particles from gen-1, gen-2, and gen-3, respectively. The red line in (c1) shows the normalized laser field intensity. The laser and electron-beam parameters are given in the main text.}
\end{figure}

The electron bunch counter-propagates along the $-z$ direction with an initial energy of $10~\mathrm{GeV}$. It has an angular divergence of $\Delta\theta=0.1~\mathrm{mrad}$, a relative energy spread of $\Delta\varepsilon_0/\varepsilon_0=0.01$, a transverse radius of $w_e=0.4~\mu\mathrm{m}$, a longitudinal length of $L_e=5~\mu\mathrm{m}$, and a total electron number of $N_e=5\times10^6$. The bunch has a Gaussian transverse density profile and a uniform longitudinal distribution. These electron-bunch parameters are kept fixed throughout the $(a_0,\tau)$ scan. Electron beams with comparable energies and beam qualities are accessible using conventional accelerators and laser--plasma acceleration schemes~\cite{Zhu2023,Babjak2024,Tsymbalov2025}. 
Using the maximum pair yield obtained in the representative simulation, the maximum electron or positron density is conservatively estimated as $n_{-}^{\max}\sim3.6\times10^{25}~\mathrm{m}^{-3}$. Even assuming complete transverse charge separation over $l\simeq w_e=0.4~\mu\mathrm{m}$, the resulting electrostatic field is only $E_{\mathrm{sep}}^{\max}\sim e n_-^{\max}l/\varepsilon_0\simeq2.6\times10^{11}~\mathrm{V/m}$, corresponding to $E_{\mathrm{sep}}^{\max}/E_0\lesssim4\times10^{-4}$ throughout the considered range $a_0=200$--$1000$ and confirming that plasma collective fields can be neglected under the present conditions.

The temporal and generation resolved particle yields are presented in Fig.~\ref{fig2}. 
In Fig.~\ref{fig2}(a), the photon yields exhibit non-monotonic generation dependence, peaking at gen-2 with a ratio $N_{\gamma,2}/N_{\gamma,1}\approx 10.2$ and declining at gen-3 ($\approx 6.0$), while pair yields follow the similar trend ($1.0 : 1.4 : 0.2$) [Fig.~\ref{fig2}(b)]. 
Temporally, particle production for each generation begins before the laser intensity maximum [red curve in Fig.~\ref{fig2}(c1)], with phase delays between successive generations evident in the photon [Figs.~\ref{fig2}(c1)-(c3)] and pair [Figs.~\ref{fig2}(d1)-(d3)] phase-space distributions. 
Additionally, earlier-generation particles gain higher energies from their parent particles, as reflected in the spectral distributions [Figs.~\ref{fig2}(c2,d2) vs (c3,d3)].

This hierarchy arises from the competition between particle multiplication and depletion. 
Initially, a limited number of seed particles yield modest first-generation emission [blue curves in Figs.~\ref{fig2}(a) and (c1)]. 
As hard photons accumulate, a feedback loop activates---newly created pairs emit secondary photons, amplifying the source of the cascade and enhancing  second-generation growth [orange curves in Figs.~\ref{fig2}(a), (c2), and (d2)]. 
Subsequently, radiation reaction (RR) redistributes energy among proliferating secondaries, softens the photon spectrum, and further suppresses third-generation production via enhanced depletion [lavender curves in Figs.~\ref{fig2}(a), (c3), and (d3)].

\begin{figure}
    \centering
    \begin{overpic}{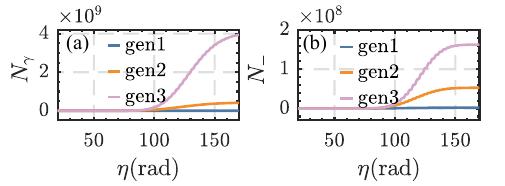}
    \end{overpic}
    \caption{\label{fig3}
Generation-resolved cumulative particle yields as functions of $\eta$ in the absence of radiation reaction (RR) effect. The left and right panels correspond to the photon yield $N_\gamma$ and electron yield $N_-$, respectively. The blue, orange, and lavender curves represent the cumulative yields up to gen-1, gen-2, and gen-3, respectively. The parameters are all the same as those in Fig.~\ref{fig2}.} 
\end{figure}

To examine the impact of RR, Fig.~\ref{fig3} presents the cumulative photon and particle yields versus laser phase $\eta$ without RR. 
Comparing the cases including and excluding RR, the first-generation yields remain nearly unchanged, whereas the gen-$2$ and gen-$3$ photon yields with RR are reduced to $11.75\%$ and $1.37\%$, respectively, of the corresponding values without RR [Fig.~\ref{fig3}(a)]. The electron yields exhibit the same trend [Fig.~\ref{fig3}(b)].
 
Consequently, the cascade hierarchy inverts: gen-2 dominates the total yield under RR, whereas gen-3 prevails in the excluding RR case. 
Thus, radiation reaction not only reduces the overall particle production but also truncates the cascade at earlier stages.

\begin{figure}
    \centering
    \begin{overpic}{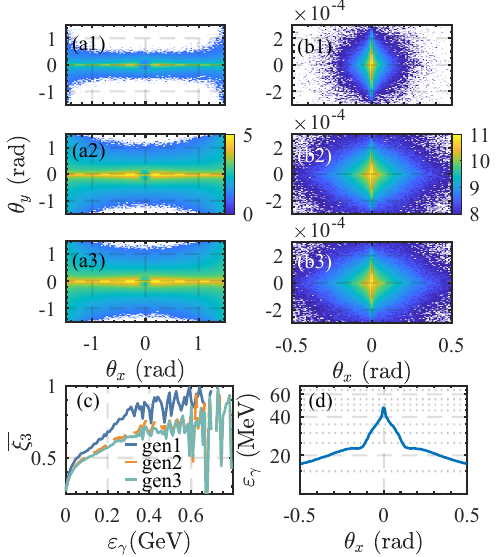}
    \end{overpic}
    \caption{\label{fig4}(a1) -- (a3) Angle-resolved accumulative photon density $\log_{10}[\mathrm{d}^2N_\gamma/(\mathrm{d}\theta_x\mathrm{d}\theta_y)]$ over $\theta_x$ and $\theta_y$. (a1) the first generation, (a2) the second generation and the first generation, (a3) the third generation and the previous generations. (b1) -- (b3) Angle resolved accumulative electron density $\log_{10}[\mathrm{d}^2N_{-}/(\mathrm{d}\theta_x\mathrm{d}\theta_y)]$ over $\theta_x$ and $\theta_y$. (b1) the first generation, (b2) the second generation and the first generation, (b3) the third generation and the previous generations. (c) The average photon polarization $\overline{\xi}_3$ over photon energy $\varepsilon_\gamma$, blue solid line the first generation, orange dashed line the second generation and the first generation, and green solid line the third generation and the previous generations. (d) Photon energy $\varepsilon_{\gamma}$ over $\theta_x$. The parameters are the same as those in Fig.~\ref{fig2}.}
\end{figure}

Fig.~\ref{fig4} presents the angle-resolved particle distributions across generations, revealing the cumulative angular broadening and evolution of the photon and pair populations. Gen-$1$ photons [Fig.~\ref{fig4}(a1)] exhibit an initial angular spread consistent with the characteristic emission angle $\theta_0 \approx a_0/\gamma_e \approx 0.03$~rad for $\gamma_e \approx 2\times10^4$ and $a_0=600$, whereas the corresponding pairs [Fig.~\ref{fig4}(b1)] remain strongly collimated. This difference results from the energy- and angle-dependent NBW conversion probability. Photons emitted at larger angles generally have lower energies [Fig.~\ref{fig4}(d)] and propagate less directly against the laser field. Both effects reduce their quantum parameter $\chi_\gamma$ and suppress pair production. Consequently, the generated pairs predominantly inherit the momenta of the high-energy photons emitted close to the electron propagation direction, leading to a narrower angular distribution than that of the parent photon population.
As the cascade develops to the second generation, stochastic emission and RR progressively broaden the photon angular distribution [Fig.~\ref{fig4}(a2)], whereas pair production [Fig.~\ref{fig4}(b2)] maintains its narrow profile. By gen-$3$ [Figs.~\ref{fig4}(a3) and (b3)], the substantially reduced yields of both photons and electrons result in angular distributions that closely resemble those of gen-2, indicating saturation of the broadening effect. 
Additionally, the ensemble-averaged photon Stokes parameter $\overline{\xi}_3$ decreases with cascade generation [Fig.~\ref{fig4}(c)]. The seed electrons initially propagate nearly along the $-z$ direction, so the photons they emit share a relatively uniform emission geometry and linear-polarization orientation. In higher generations, however, the secondary electrons and positrons have broader momentum directions and are no longer confined to the $-z$ axis. The local polarization bases of their emitted photons therefore vary across the ensemble. When the photon polarization states are expressed in a common basis, this directional spread reduces the net $\xi_3$ component, resulting in the observed decrease of $\bar{\xi}_3$ with generation.

\section{The impact of laser parameters on the QED cascade}

\begin{figure}[!h]
    \centering
    \begin{overpic}{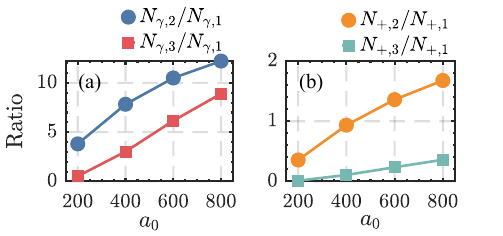}
    \end{overpic}
    \caption{Generation-resolved yield ratios versus the laser amplitude $a_0$.
(a) Photon-yield ratios $N_{\gamma,2}/N_{\gamma,1}$ and $N_{\gamma,3}/N_{\gamma,1}$ as functions of $a_0$.
(b) Positron-yield ratios $N_{+,2}/N_{+,1}$ and $N_{+,3}/N_{+,1}$ as functions of $a_0$.
Here $N_{\gamma,1(2,3)}$ and $N_{+,1(2,3)}$ denote the total yields of the first, second, and third generations for photons and positrons, respectively.
All other parameters are the same as those in Fig.~\ref{fig2}.}
    \label{fig5}
\end{figure}

Fig.~\ref{fig5} shows the dependence of the generation-yield ratios on the laser amplitude $a_0$. For both photons and positrons, the gen-2/gen-1 and gen-3/gen-1 ratios increase monotonically with $a_0$ [Figs.~\ref{fig5}(a) and (b)]. Throughout the investigated range, the gen-2/gen-1 ratio remains larger than the corresponding gen-3/gen-1 ratio. The separation between the two ratios is more pronounced for positrons, whereas the photon ratios exhibit a comparatively smaller difference. These results show that the relative contribution of higher generations increases with laser amplitude.

\begin{figure}[!h]
    \centering
    \begin{overpic}{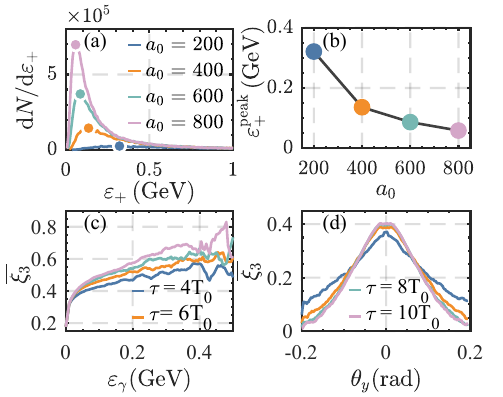}
    \end{overpic}
    \caption{\label{fig6}(a) Positron energy spectra d$N/\mathrm{d}\varepsilon_{+}(\mathrm{GeV}^{-1})$ for $a_0 = 200, 400, 600$, and $800$ (represented by blue, orange, green, and lavender lines respectively). The dots in (a) indicate the peak energies $\varepsilon_+^\mathrm{peak}$ in the spectra and are replotted in (b) with respect to the laser peak intensity $a_0$. (c) and (d) The average polarization degree of photons $\overline{\xi}_3$ for different laser pulse durations $\tau$, with respect to the photon energy $\varepsilon_\gamma$~(GeV) and photon angle $\theta_y$. Here, the blue, red, green, and black lines indicate $\tau = 4T_0, 6T_0, 8T_0$, and $10T_0$, respectively.
    All other parameters are the same as those in Fig.~\ref{fig2}.}
\end{figure}

\begin{figure}
    \centering
    \begin{overpic}{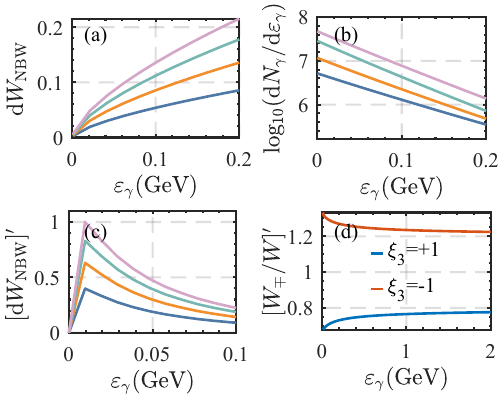}
    \end{overpic}
    \caption{\label{fig7}(a) Normalized probability of pair production $\mathrm{d}W_{\mathrm{NBW}}(\mathrm{GeV}^{-1})$ versus photon energy $\varepsilon_\gamma\mathrm{(GeV)}$ at  different laser intensities $a_0$, blue line $a_0 = 200$, orange line $a_0 = 400$, green line $a_0 = 600$, lavender line $a_0 = 800$. (b) The energy density  of photons  $\log_{10}(\mathrm{d}N_\gamma/\mathrm{d}\varepsilon_\gamma)$ emitted at different laser intensities $a_0$ via cascades versus photon energy $\varepsilon_\gamma\mathrm{(GeV)}$. (c) Photon energy multiplied by probability of laser intensity $\mathrm{d}N_\gamma/\mathrm{d}\varepsilon_{\gamma}\cdot\mathrm{d}W_{\mathrm{NBW}}$ versus photon energy $\varepsilon_\gamma\mathrm{(GeV)}$. (d) The ratio of the pair production probability for fully polarized photons to that of unpolarized photons as a function of photon energy $\varepsilon_\gamma$. The curves correspond to the Stokes parameters $\xi_3 = +1$ (blue) and $\xi_3 = -1$ (red), normalized by the unpolarized baseline ($\xi_3 = 0$). The legends of (a), (b), and (c) are the same as those in Fig.~\ref{fig6}(a). The parameters here are the same as those in Fig.~\ref{fig2}.}
\end{figure}

Fig.~\ref{fig6}(a) shows that increasing $a_0$ broadens the positron energy spectrum and shifts its peak, $\varepsilon_+^{\rm peak}$, toward lower energies, as summarized in [Fig.~\ref{fig6}(b)]. This spectral softening results from the stronger radiative energy loss of the primary and secondary electrons at larger $a_0$. 
The average photon polarization $\overline{\xi}_3$ with respect to photon energy $\varepsilon_\gamma$ exhibits a clear positive dependence on the pulse duration [see Fig.~\ref{fig6}(c)].
As shown in Fig.~\ref{fig6}(d), the angular dependencies of $\overline{\xi}_3$ nearly overlap for the cases of $\tau = 6T_0, 8T_0$, and $10 T_0$. In Fig.~\ref{fig7} a detailed analysis about above results is presented. 

The left-shifting of $\varepsilon_{+}^{\rm peak}$ arises from the product of the photon spectrum with the energy-resolved pair-production rate, $\mathrm{d}N_+/\mathrm{d}\varepsilon_+ \propto (\mathrm{d}N_\gamma/\mathrm{d}\varepsilon_\gamma) \times (\mathrm{d}W_{\rm NBW}/\mathrm{d}\varepsilon_+)$. 
Fig.~\ref{fig7}(a) shows that the NBW probability increases with both photon energy $\varepsilon_\gamma$ and laser intensity $a_0$, while [Fig.~\ref{fig7}(b)] reveals that the photon spectrum itself exhibits an approximately exponential decay. 
In Fig.~\ref{fig7}(c), the product of these two trends yields peaks at progressively lower energies as $a_0$ increases, explaining the monotonic shift of $\varepsilon_{+}^{\rm peak}$ observed in [Fig.~\ref{fig6}(b)].

The increase of $\overline{\xi}_3$ with pulse duration can be attributed to polarization-selective depletion during NBW pair production. As shown in [Fig.~\ref{fig7}(d)], under the present Stokes-parameter convention, photons with $\xi_3=-1$ have a higher
pair-production probability than those with $\xi_3=+1$. The $\xi_3<0$ component is therefore preferentially converted into pairs, biasing the surviving photon population toward positive
values of $\xi_3$. Increasing $\tau$ extends the interval over which this filtering mechanism operates and allows its effect to accumulate through successive cascade steps, leading to the larger values of $\overline{\xi}_3$ observed in [Figs.~\ref{fig6}(c) and \ref{fig6}(d)], consistent with Ref.~\cite{Lv2025}.


\section{Laser-Parameter Diagnostics from Cascade Signatures}\label{sec4}
As mentioned in the previous section, we have discovered clear dependencies between the peak of the positron energy spectrum and the laser field strength $a_0$, and correlations between the average photon polarization and the duration of the laser pulse.
Therefore, we propose the diagnostic method based on QED cascade to infer the laser amplitude $a_0$ and pulse duration $\tau$.

The quantities considered here characterize different aspects of the laser--particle interaction.
The maximum cascade generation $G_{\max}$ is a model-defined measure of the cascade depth, whereas the fraction of backward-emitted photons $F_r$ and the positron peak energy $\varepsilon_+^{\rm peak}$ are experimentally accessible final-state observables.

We systematically vary the laser amplitude from $a_0=200$ to $1000$ in increments of 50
and the pulse duration from $\tau=2T_0$ to $12T_0$ in increments of $0.5T_0$, while keeping other laser parameters and the electron-bunch parameters specified in Sec.~\ref{sec3} fixed. The resulting $17\times21$ parameter grid contains 357 simulation cases.

\begin{figure}
    \centering
    \begin{overpic}[width=\linewidth]{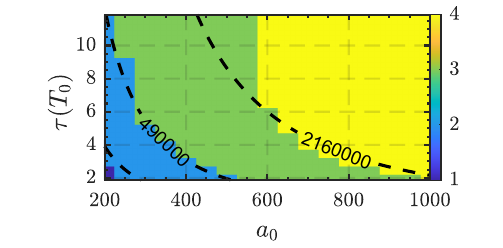}
    \end{overpic}
    \caption{
    Heatmap of the maximum cascade generation number $G_{\max}$ in the $(a_0,\tau)$ parameter space. The black contours indicate iso-values of $a_0^{2}\tau$ (proportional to the laser energy) and are overlaid to guide comparison across parameter sets.
    All other parameters are the same as those in Fig.~\ref{fig2}.
    }
    \label{fig8}
\end{figure}

Fig.~\ref{fig8} shows the maximum cascade generation number $G_{\max}$ as a function of $a_0$ and $\tau$. Here, $G_{\max}$ is defined as the highest generation whose particle yield remains above the termination threshold specified in Sec.~\ref{sec2}. The black curves are iso-contours of $a_0^2\tau$, which is proportional to the total laser-pulse energy, $\mathcal{E}_L\propto I\tau w_0^2\propto a_0^2\tau$, for a fixed focal waist $w_0$. The distribution of $G_{\max}$ closely follows these contours, indicating that the cascade generation depth is determined primarily by the total laser energy deposited within the interaction volume. From the distribution of $G_{\max}$ in the $(a_0,\tau)$ parameter space, one can directly estimate the cascade generation reached in a given laser--electron collision experiment. For example, the parameter set ($a_0 = 300, \tau  = 6T_0$), located in the green region corresponds to cascades developing up to the third generation ($G_{\max}=3$).

The approximate correlation between $G_{\max}$ and $a_0^2\tau$ can be understood qualitatively from the accumulated nature of cascade development.
Increasing $a_0$ enhances the NCS and NBW probabilities, whereas increasing $\tau$ extends
the interval over which successive emission and pair-production steps can occur.
Both effects increase the likelihood that the cascade reaches a higher generation before the particles leave the focal region or their energies fall below those required for further multiplication.

The correspondence with the $a_0^2\tau$ contours is not expected to be exact.
At low $a_0$ or short $\tau$, pair production is strongly limited by the photon-energy threshold and the exponential sensitivity of the NBW rate to $\chi_\gamma$.
At large $a_0$ and long $\tau$, radiative energy depletion reduces the energy available to successive generations, causing the increase of $G_{\max}$ to become progressively weaker.

 \begin{figure}
    \centering
    \begin{overpic}[width=\linewidth]{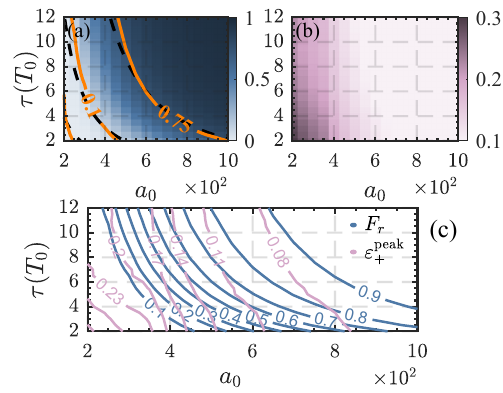}
    \end{overpic}
    \caption{
    (a) Heatmap of the fraction of backward-emitted photons, $F_r = N_\gamma(p_z>0)/N_\gamma^{\rm total}$, mapped in $(a_0,\tau)$ space. The black $a_0^{2}\tau$ iso-contours are shown to highlight the correlation between photon momentum reversal and laser intensity. The orange curves are iso-contours of $F_r$.
    (b) Heatmap of the peak energy of positrons, $\varepsilon_+^\mathrm{peak}\mathrm{(GeV)}$, mapped in the same $(a_0,\tau)$ space.
    (c) Contour lines extracted from panels~(a) and (b). The blue curves indicate iso-values of $F_r$, while the lavender curves indicate iso-values of $\varepsilon_{+}^{\rm peak}$. The intersections between the two sets of contour lines can be used to diagnose the laser parameters simultaneously.
    All other parameters are the same as those in Fig.~\ref{fig2}.
    }
    \label{fig9}
\end{figure}

Fig.~\ref{fig9}(a) shows the fraction of backward-emitted photons
\begin{equation}
F_r=\frac{N_\gamma(p_z>0)}{N_\gamma^{\rm total}}
\end{equation}
in the $(a_0,\tau)$ parameter space.
Since the incident electron bunch propagates along the $-z$ direction, the condition $p_z>0$ identifies photons emitted opposite to the initial electron-beam direction.

The iso-contours of $F_r$ approximately follow those of $a_0^2\tau$, indicating that $F_r$ is sensitive to the accumulated strength of the laser--particle interaction.
A larger $a_0$ enhances radiative energy loss, whereas a longer pulse allows repeated emission and recoil to accumulate over a longer interaction interval.
These effects increase the probability that charged particles reverse their longitudinal motion in the laser field and subsequently emit photons with $p_z>0$.
Consequently, $F_r$ generally increases as the cascade develops more deeply.
The correlation between $F_r$ and $G_{\max}$ suggests that the experimentally accessible quantity $F_r$ can serve as an indirect proxy for the model-defined cascade depth.

Fig.~\ref{fig9}(b) presents the peak energy of the produced positrons, $\varepsilon_{+}^{\rm peak}\mathrm{(GeV)}$, in the same parameter space. In contrast to $F_r$, which mainly follows the accumulated cascade strength and exhibits an approximate dependence on the combined scaling $a_0^2\tau$, $\varepsilon_{+}^{\rm peak}$ is more directly governed by the instantaneous field strength, and hence is primarily sensitive to the laser amplitude $a_0$. A larger $a_0$ increases the local quantum parameters of both radiating electrons and emitted photons, thereby modifying the typical photon energy and the subsequent energy partition in NBW pair creation. As a result, $\varepsilon_{+}^{\rm peak}$ provides complementary information to $F_r$. 

As illustrated in [Fig.~\ref{fig9}(c)], a measured pair $\left(F_r,\varepsilon_+^{\rm peak}\right)$ selects one iso-contour from each observable map.
The intersections of the two contours provide candidate solutions for the laser parameters $(a_0,\tau)$.
Over most of the scanned parameter space, the two contour families remain sufficiently nonparallel to permit a locally unique reconstruction.
The corresponding sensitivity to measurement uncertainties is quantified below using the local Jacobian of the observable map.

\begin{figure}
    \centering
    \begin{overpic}[width=\linewidth]{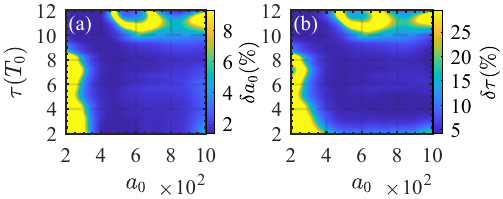}
    \end{overpic}
    \caption{
    (a) Relative error of the diagnosed laser amplitude $a_0$ obtained from the contour-line intersections of $F_r$ and $\varepsilon_{+}^{\rm peak}$ in the parameter space.
    (b) Relative error of the diagnosed pulse duration $\tau$ obtained from the contour-line intersections of $F_r$ and $\varepsilon_{+}^{\rm peak}$ in the parameter space.
    }
    \label{fig10}
\end{figure}

To quantify the local precision of the reconstruction, we propagate the measurement uncertainties of $F_r$ and $\varepsilon_+^{\rm peak}$ to the inferred laser parameters.
Let the observable and parameter vectors be
\begin{equation}
\bm{O}
=
\begin{pmatrix}
F_r\\
\varepsilon_+^{\rm peak}
\end{pmatrix},
\qquad
\bm{p}
=
\begin{pmatrix}
a_0\\
\tau
\end{pmatrix}.
\end{equation}
Near a reconstructed point
$\bm{p}^{*}=(a_0^{*},\tau^{*})$, small variations in the
observables and laser parameters are related through the local
Jacobian matrix,
\begin{equation}
\Delta\bm{O}
=
J(\bm{p}^{*})\,\Delta\bm{p},
\qquad
J_{ij}(\bm{p}^{*})
=
\left.
\frac{\partial O_i}{\partial p_j}
\right|_{\bm{p}=\bm{p}^{*}} .
\end{equation}
Let $C_O$ denote the covariance matrix of the measured observables.
The covariance matrix of the reconstructed laser parameters is then obtained by linear error propagation as
\begin{equation}
C_p
=
J^{-1}(\bm{p}^{*})\,
C_O\,
\left[J^{-1}(\bm{p}^{*})\right]^{T}.
\end{equation}
Assuming that the measurement uncertainties of $F_r$ and $\varepsilon_+^{\rm peak}$ are uncorrelated, the observable covariance matrix is
\begin{equation}
C_O
=
\begin{pmatrix}
\sigma_F^2 & 0\\
0 & \sigma_E^2
\end{pmatrix},
\end{equation}
where $\sigma_F$ and $\sigma_E$ denote the standard uncertainties of $F_r$ and $\varepsilon_+^{\rm peak}$, respectively.
The marginal standard uncertainties of the reconstructed laser amplitude and pulse duration are obtained from the diagonal elements of $C_p$:
\begin{equation}
\sigma_{a_0}
=
\sqrt{(C_p)_{11}},
\qquad
\sigma_{\tau}
=
\sqrt{(C_p)_{22}}.
\end{equation}
The corresponding relative uncertainties are defined as
\begin{equation}
\delta_{a_0}
=
\frac{\sigma_{a_0}}{a_0^{*}},
\qquad
\delta_{\tau}
=
\frac{\sigma_{\tau}}{\tau^{*}}.
\end{equation}

To provide an illustrative estimate of the local reconstruction precision, we assume independent standard uncertainties of $\sigma_F=0.01$ for the photon momentum-reversal fraction and
$\sigma_E=0.01~\mathrm{GeV}$ for the positron spectral-peak energy.
The latter corresponds to a few-percent energy uncertainty for positron peaks in the several-hundred-MeV range, consistent with the percent-level spectral resolution anticipated for magnetic
electron--positron spectrometers in strong-field QED experiments~\cite{Borysov2022,Salgado2022}. The assumed values should be regarded as benchmark measurement uncertainties rather than as performance specifications of a particular experimental setup; a facility-specific analysis would
require the detector acceptance, response, background, and calibration uncertainties to be included explicitly.
For a counting-dominated measurement, $\sigma_F=0.01$ requires approximately $10^3$--$10^{3.5}$
detected photons for the range of $F_r$ considered here, although systematic uncertainties associated with the relative acceptance and efficiency of the two photon detectors may dominate in practice.

Using the local Jacobian at each point of the parameter grid, we calculate the relative uncertainties $\delta_{a_0}$ and $\delta_{\tau}$ shown in Figs.~\ref{fig10}(a) and \ref{fig10}(b), respectively. The reconstruction is most precise in regions where $F_r$ and $\varepsilon_+^{\rm peak}$ vary appreciably with the laser parameters and their gradients are sufficiently distinct, yielding a well-conditioned Jacobian. Small gradients or nearly parallel contour families instead amplify the propagated uncertainties. For the present observable map and assumed measurement errors, the lowest uncertainties occur mainly in the intermediate-to-high-$a_0$ region, demonstrating that the joint measurement of $F_r$ and $\varepsilon_+^{\rm peak}$ can provide simultaneous local constraints on $a_0$ and $\tau$.

\section{Conclusion}

In this work, we have presented a generation-resolved, spin- and polarization-dependent analysis of shower-type QED cascades produced in head-on laser--electron collisions over the ranges $a_0=200$--$1000$ and $\tau=2T_0$--$12T_0$. The final-state photon and positron distributions retain clear information about the laser parameters, enabling a post-interaction reconstruction based on the backward-photon fraction $F_r$ and the positron spectral-peak energy $\varepsilon_+^{\rm peak}$. While $F_r$ mainly reflects the accumulated cascade evolution and is approximately organized by $a_0^2\tau$, $\varepsilon_+^{\rm peak}$ is more directly sensitive to $a_0$; their distinct parameter dependences therefore provide simultaneous constraints on the laser amplitude and pulse duration through simulation-based observable maps. The maximum cascade generation $G_{\max}$ and the average photon polarization $\overline{\xi}_3$ further characterize the cascade depth and polarization-selective pair-production dynamics. More generally, the generation-resolved framework separates the contributions of successive generations to the particle spectra, angular distributions, and polarization signals, providing a useful tool for studying cascade kinetics and diagnosing laser parameters in future strong-field QED experiments, with potential applications to laser-driven nuclear physics, laboratory astrophysics, and searches for physics beyond the Standard Model.

{\it Acknowledgments---}The work is supported by the National Natural Science Foundation of China (Grants No. 12425510, No. U2267204, No. 12441506, and 12475249), the National Key Research and Development (R\&D) Program (Grant No. 2024YFA1610900 and 2024YFA1612700), the Science Challenge Project (No. TZ2025012), and the Innovative Scientific Program of CNNC.

\bibliography{aipsamp}
\end{document}